\def\year{2022}\relax
\documentclass[letterpaper]{article} 
\usepackage{aaai22}  
\usepackage{times}  
\usepackage{helvet}  
\usepackage{courier}  
\usepackage[hyphens]{url}  
\usepackage{graphicx} 
\usepackage{natbib}  
\usepackage{caption} 
\DeclareCaptionStyle{ruled}{labelfont=normalfont,labelsep=colon,strut=off} 
\usepackage{algorithm}
\usepackage{algorithmic}
\usepackage{amsfonts}
\usepackage{amsmath}
\usepackage{booktabs}
\usepackage{newfloat}
\usepackage{listings}
\floatstyle{ruled}
\newfloat{listing}{tb}{lst}{}
\floatname{listing}{Listing}

\usepackage{multirow}

\title{Fair Conformal Predictors for Applications in Medical Imaging}
\author {
    Authors 
}
\author {
    Charles Lu,\textsuperscript{\rm 1, $\dagger$}
    Andr\'{e}anne Lemay,\textsuperscript{\rm 2}
    Ken Chang,\textsuperscript{\rm 3, 4}
    Katharina H\"{o}bel,\textsuperscript{\rm 3, 4}
    Jayashree Kalpathy-Cramer\textsuperscript{\rm 1, 4}
}
\affiliations {
    \textsuperscript{\rm 1} Athinoula A. Martinos Center for Biomedical Imaging\\
    \textsuperscript{\rm 2} Polytechnique Montr\'{e}al\\
    \textsuperscript{\rm 3} Massachusetts Institute of Technology\\
    \textsuperscript{\rm 4} Harvard Medical School\\
    \textsuperscript{$\dagger$} clu@mgh.harvard.edu\\
}

\begin{document}
\maketitle

\begin{abstract}
    Deep learning has the potential to automate many clinically useful tasks in medical imaging. 
    However translation of deep learning into clinical practice has been hindered by issues such as lack of the transparency and interpretability in these ``black box'' algorithms compared to traditional statistical methods.
    Specifically, many clinical deep learning models lack rigorous and robust techniques for conveying certainty (or lack thereof) in their predictions -- ultimately limiting their appeal for extensive use in medical decision-making.
    Furthermore, numerous demonstrations of algorithmic bias have increased hesitancy towards deployment of deep learning for clinical applications.
    To this end, we explore how conformal predictions can complement existing deep learning approaches by providing an intuitive way of expressing uncertainty while facilitating greater transparency to clinical users.
    In this paper, we conduct field interviews with radiologists to assess possible use-cases for conformal predictors. 
    Using insights gathered from these interviews, we devise two clinical use-cases and empirically evaluate several methods of conformal predictions on a dermatology photography dataset for skin lesion classification. 
    We show how to modify conformal predictions to be more adaptive to subgroup differences in patient skin tones through equalized coverage.
    Finally, we compare conformal prediction against measures of epistemic uncertainty.
\end{abstract}

\section{Introduction}
    As the cost of healthcare continues to rise in many countries around the world, massive investment has been poured into medical AI development in the hopes of leveraging these technologies to lower costs and improve efficiency~\cite{ai-index, Topol2019HighperformanceMT}.
    However translation into clinical deployment has proved extremely difficult -- exemplified by several highly publicized projects ultimately failing to be widely used in clinical practice~\cite{ibmfail}.
    
    Medical AI tools can pose unique design considerations that significantly differ from more traditional medical device software~\cite{DBLP:journals/corr/abs-2109-06919}. 
    Regulatory agencies such as the U.S. Food and Drug Administration (FDA) have begun to consider these differences with more scrutiny~\cite{fda}. 
    Research communities such as human-computer interaction also increasingly explore the unique challenges of designing machine learning for healthcare applications~\cite{JacobsHPLAMPDG21,Xie2020CheXplainEP,10.1145/3313831.3376718}. 
    
    One challenging obstacle for clinical machine learning is identifying cases in which the model has low confidence -- a scenario that might necessitate further review of those cases. 
    Furthermore, the lack of transparency in machine learning models has also been cited as a major barrier to developing AI tools that are adopted by clinicians~\cite{Yang2019UnremarkableAF,Begoli2019TheNF}.
    A survey by~\citet{Allen20212020AD} showed that even though many radiologists believed that AI tools added value to their clinical practice, they would not trust AI for autonomous clinical use.
    
    Deep learning models, in particular, are notoriously difficult to interrogate and known to be vulnerable to a number of attacks that can compromise the safety and privacy of patients~\cite{doi:10.1148/ryai.2021200267, doi:10.1126/science.aaw4399}.
    Also, many recent studies highlight the risks of algorithmic biases in machine learning for applications such as skin lesion classification and facial recognition systems~\cite{bissoto19deconstructing,pmlr-v81-buolamwini18a,pierson2021algorithmic,banerjee2021reading}.
    
    To address these challenges, we explore the use of conformal predictions methods for medical imaging applications.
    We argue that conformal predictions may better correspond to clinical decision-making intuition and can be easily incorporated into existing models while providing meaningful statistical guarantees~\cite{10.1007/978-3-030-50146-4_39}. 
    For example, doctors routinely express uncertainty in the form of comparative sets to arrive at a diagnosis i.e. a \textit{differential diagnosis}.
    Additionally, conformal prediction sets naturally convey a measure of uncertainty by the number of items contained in their set.
    Several works promote uncertainty quantification to facilitate greater trust in medical black-box models and to detect bias in protected patient demographics~\cite{10.1145/3461702.3462571,kompa,lu2021evaluating}.
    
    In this study, we demonstrate the potential of conformal predictors to offer a more clinically intuitive representation of model uncertainty.
    Based on interviews with clinicians, we characterize two conformal use-cases (\textit{rule-in} and \textit{rule-out}) and show how conformal predictors can be adapted to yield equalized coverage at a subgroup level.
    We empirically evaluate our group conformal predictors on a dermatology dataset for skin lesion classification using Fitzpatrick skin type as our group attribute.
    Finally, we compare conformal uncertainty against epistemic uncertainty to show how group-calibrated conformal predictors better represent relevant subgroup differences such as disease prevalence of malignant skin conditions.
        
\section{Related Work}
    \subsection{Fairness in Machine Learning}
        A major challenge for the deployment of machine learning in healthcare is the potential to encode and and reinforce inequalities among protected patient demographics~\cite{Larrazabal12592,Pierson2021AnAA}.
        Attempts have been made to formally conceptualize ``fairness'' statistically and definitions of fairness have been proposed to measure disparity between demographic groups such as race and gender~\cite{barocas-hardt-narayanan}.
        
        Examples of group fairness metrics for binary classification, $Y \in \{0, 1\}$, include \textit{demographic parity}:
        \begin{equation}
            \small
            P(\hat{Y} = 1 \mid A = a) \; = \; P(\hat{Y} = 1 \mid A = b), 
        \end{equation}
        which desires independence between the group attribute, $a, b \in A$, and the predicted response $\hat{Y}$, \textit{equalized odds}:
        \begin{equation}
            \small
            P(\hat{Y} = 1 \mid Y = y, A = a) \; = \; P(\hat{Y} = 1 \mid Y = y, A = b),
        \end{equation}
        for $y \in \{0, 1\}$ and desires conditional independence, $A \bot \hat{Y} \mid Y$, and \textit{calibration parity}, which is achieved only if the prediction scores, $S \in [0, 1]$, for each subgroup are perfectly calibrated
        \begin{equation}
            \small
            s = P(Y = y \mid S = s, A = a) = P(Y = y \mid S = s, A = b)
        \end{equation}
        for all score thresholds, $\forall s \in S$.
        
        As shown in \citet{kleinberg_et_al:LIPIcs:2017:8156}, not all fairness metrics are mutually satisfiable, and many metrics assume the observed outcome for all subgroups is known and measurable (an assumption that may not be practical in many clinical situations).
        Furthermore, many bias mitigation strategies often assume access to the model's training procedure or make explicit distributional assumptions~\cite{Zafar_2017, NIPS2017_a486cd07}. 
        For regulated medical device software, these assumptions may not be realistic as medical device manufacturers would not permit such access or modifications to their proprietary, commercial models.
        
    \subsection{Uncertainty for Deep Learning}
        \textit{Epistemic uncertainty} concerns the probabilistic estimation of model parameters. 
        Several approaches to uncertainty have been developed for deep neural networks, such as Bayesian deep learning, which encompasses different approaches to approximating full Bayesian neural networks~\cite{Gal2016Uncertainty}. 
        Techniques such as Monte-Carlo dropout modify the architecture or training procedure of the model to learn a posterior distribution over the parameters~\cite{pmlr-v48-gal16}.
        For example, an ensemble of deep models can be trained with adversarial examples and then averaged to estimate the posterior distribution~\cite{NIPS2017_9ef2ed4b}.
        
        Regardless of which technique is used, the posterior distribution can be randomly sampled to estimate different measures of epistemic uncertainty, such as \textit{predictive entropy}, which is the average entropy over $T$ samples and $K$ classes,
        \begin{equation}
            \label{eq:entropy}
            \small
                -\frac{1}{K} \sum_{k=1}^K \left(\frac{1}{T} \sum_{t=1}^T p_{t \mid k} \log \frac{1}{T} \sum_{t=1}^T p_{t \mid k}\right).
        \end{equation}
        
        
        An alternative approach to uncertainty involves calibrating model predictions such that the confidence of a prediction matches the actual probability that that prediction would be correct for all confidence levels. 
        Deep neural networks are generally regarded to be poorly calibrated.
        This has motivated many post-hoc methods, such as Platt Scaling, to help better calibrate the outputs of the softmax function~\cite{pmlr-v70-guo17a}:
        \begin{equation}
            \label{eq:platt}
            \small
            \textrm{Platt}(z) = \frac{e^{z/\beta}}{\sum_i e^{z_i/\beta}},
        \end{equation}
        where $z$ is the uncalibrated output and $\beta$ is a learned parameter that minimizes negative log-likelihood on the validation set.
        
    \subsection{Conformal Predictions}
    
        One limitation of epistemic uncertainty is its lack of any standard or intuitive meaning to aid users in their decision making.
        While post-hoc calibration techniques do have a statistical interpretation, they do not usually provide any distribution-free guarantees of reliability or miscalibration.
        Additionally, common approaches often require access to the model (e.g. training with a modified loss function) or distributional assumptions (e.g. sampling from a Bernoulli distribution), which may not be feasible in third-party medical device software.
        Conformal prediction methods do not have these limitations.
        
        First introduced by \citet{vovk}, conformal prediction is a method of distribution-free uncertainty quantification, which is model agnostic and provides formal results for \textit{marginal coverage}, defined as the average probability that the true class will be contained in the prediction set.
        Formally, this coverage guarantees states:
        \begin{equation}
            \small
            1 - \alpha \; \leq \; P\left(y_n \; \in \; \left\{y \in \mathcal{Y} \mid S(x_n)_y > \hat{q} \right\} \right),
        \end{equation}
        where $\alpha$ is some predetermined miscoverage level, $y_n$ is the true class, $x_n$ is the unseen test example, $S(x_n)_y$ is the $y$th index of the output score, and $\hat{q}$ is the estimated score quantile needed to achieve proper coverage at the desired level.
        The score quantile is estimated on a held-out set (assumed to be IID with the test set) to ensure marginal coverage for any classifier over the joint distribution, $D_{X, Y}$.
        
        
        Any classifier that learns a score function, such as a deep neural network, can be adapted to output conformal predictions.
        For example, instead of outputting the predicted class with the highest score, e.g. $($``basal cell carcinoma''$, 0.51)$, a conformal prediction with a miscoverage level of 10\%, $\alpha=0.1$, might produce a set of predictions, e.g. $\{($``basal cell carcinoma''$, 0.51), ($ ``squamous cell carcinoma''$, 0.27), ($ ``seborrheic keratosis''$, 0.13)\}$, which will contain the true class on average 90\% of the time.
        
        These prediction sets naturally lends themselves to be utilized as a form of uncertainty and are well-suited for medical decision making~\cite{kompa,10.1007/s41666-021-00113-8}.
        Although we focus on multi-class classification, the conformal framework is general and has been extended to other prediction tasks such as regression and multi-label classification~\cite{NEURIPS2019_5103c358,JMLR:v22:20-753}.

\section{Field Interviews}
    \begin{table}[t]
    \small
    \centering
    \begin{tabular}{ccc}
        \toprule
        \textbf{participant} & \textbf{role} & \textbf{experience (years)} \\
        \midrule
        P1 & neuroradiologist & 20 \\
        P2 & radiologist & 15 \\
        P3 & neurologist & 10 \\
        P4 & clinical fellow & 5 \\
        \bottomrule
    \end{tabular}
    \caption{Role and years of experience with medical imaging of the four clinicians in our field interviews.}
    \label{tab:field}
    \end{table}
    
    To better investigate how conformal predictors could be utilized in clinical workflows, we conduct semi-structured interviews with four clinicians from a large hospital system (their role and level of experience is shown in Table~\ref{tab:field}).
    By asking participants to characterize their clinical workflow, we observed a distinction being made in the motivating reason for the imaging study, with one radiologist (P1) commenting:
    \begin{quote}
        I got to [...] decide whether it's normal or just typical expected abnormalities for age and describe those, [...] eventually there will be AI algorithms that are setup to evaluate all those things and sometimes your sensitivity and your thresholds for what you expect are different if you're just checking if everything looks normal versus if it was a study done to specifically evaluate that. 
    \end{quote}
    
    When asked how predictions of AI medical device software should be presented within the radiologist's user interface, P2 responded:
    \begin{quote}
        Don't overwhelm the radiologist with 83\% or 87\%. Is it good enough or is it not? [...] You just wanna say is this "highly confident", "cautionary", "uncertain", or "non-diagnostic"? You have to set cutoffs but you don't have to show those numbers to radiologists.
    \end{quote}
    
    In addition to wanting AI outputs to be more semantically meaningful, clinicians were also concerned about the reliability and confidence of the model's predictions.
    P1 and P3 even suggested that low confidence predictions should not be presented to the user at all depending on the clinical context.
    
    \begin{figure}[t]
    \centering
    \includegraphics[width=0.80\columnwidth]{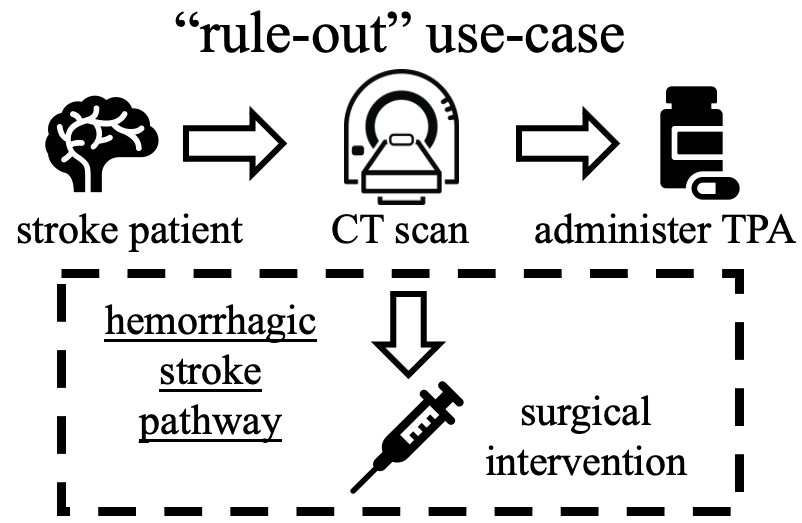} 
    \caption{A diagram of typical stroke care pathway, in which a CT exam study is ordered to ``rule-out'' hemorrhagic stroke, a contraindication for clot-breaking medication (TPA), which is the usual treatment for ischemic stroke.}
    \label{fig:rule-out}
    \end{figure}
    For example, a computed tomography (CT) imaging exam is typically performed on patients with suspected stroke to rule-out hemorrhagic bleeding, a condition that would prevent administration of tissue plasminogen activator (TPA) (Figure~\ref{fig:rule-out}).
    Thus, an application to detect hemorrhagic bleeds must prioritize sensitivity (true positive rate), even if the prediction confidence level had to be lowered, because giving TPA to a patient with a hemorrhagic stroke would worsen the bleeding. 
    This use-case of confidently ``ruling-out'' critical conditions seems to us an ideal setting for conformal prediction.
    
    \begin{figure}[t]
    \centering
    \includegraphics[width=0.80\columnwidth]{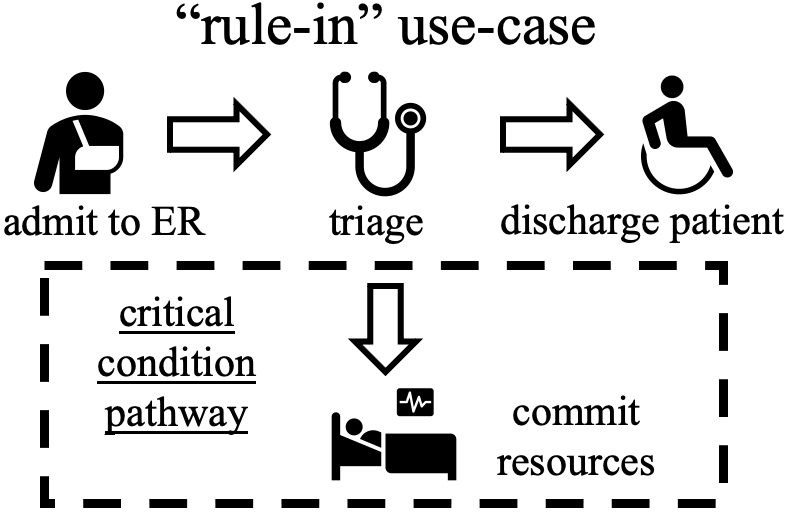} 
    \caption{A depiction of a triage workflow in the emergency room to expedite patients without severe conditions and to ``rule-in'' patients with one or more critical conditions.}
    \label{fig:rule-in}
    \end{figure}
    From our interviews, another clinical example was given: the initial triage of patients in an emergency department:
    \begin{quote}
        They want to discharge patients from the ER as quickly as possible -- healthy patients that don't need to be there. 
        Bed space is limited. 
        Sometimes they're waiting to get an answer from radiology. 
        So one of the workflows is the algorithm identifies normal cases -- no disease -- but you want it to identify patients that were definitely normal with very high confidence. 
        You want the high specificity. 
        If the doctor is overseeing 20 patients, and the algorithms said these 3 have very low chance of having any abnormality. 
        Those ones, you may feel comfortable discharging with just your look at them instead of waiting on the radiologist.
    \end{quote}
    To better allocate limited resources, patient with milder conditions may be discharged to outpatient care whereas patients more likely to have one or more severe conditions may need more immediate or invasive intervention (Figure~\ref{fig:rule-in}).
    Therefore, we formulate another use-case for conformal predictors to ``rule-in'' critical conditions.
    This ``rule-in'' use-case seems ideal for conformal predictors since the set of possible critical conditions can be very large and naturally lends itself to thinking in terms of prediction sets and possible interactions between present conditions (some conditions are of low concern when present separately but may be of higher concern if present together).
    
    Lastly, on the issue of fairness and bias, P4 saw the need for auditing platforms and monitoring dashboards to display metadata for clinically relevant demographics, such as age and race, to check for bias.
    Continual review of model predictions would be necessary to adjust decision thresholds in order to recalibrate models to changing disease prevalence over time.
    
\section{Methods}
    As conformal methods provide distribution-free coverage guarantees regardless of the number of data examples or how the machine learning model was trained, they can be easily extended to provide \textit{equalized coverage} with the group attribute is known.
    Introduced by \citet{Romano2020With} for the regression setting, equalized subgroup coverage can adapt better to differences between subgroups than aggregate conformal methods.
    
    While we acknowledge that any definition of fairness will be reductive and problematic with respect to some criteria, conformal fairness has several advantages for measuring disparity in medical AI tools. 
    First, the prevalence of many disease naturally differ between some patient demographics, such as race and age, due to various genetic, social, and behavioral risk factors.
    Attempting to encapsulate these differences solely with a statistical parity measure without the necessary clinical context would be misguided and may mislead clinical decision-making.
    Instead, outputting a set of likely predictions allows greater flexibility in clinical use-cases, such as triage and screening, by providing the clinical user more intuition to discern what should be considered normal or abnormal, on a patient level as well as a group level.
    Practically, distribution-free uncertainty quantification techniques like conformal prediction can be applied to calibrate and monitor differences between patient cohorts in arbitrary medical AI devices for real-time clinical workflows.

    We now describe a conformal algorithm to achieve equalized coverage for some coverage level, $\alpha$, on an attribute group $A$. 
    Assume that group membership is known at inference time, we can then estimate the empirical quantile, $\hat{q}_a$, for each subgroup separately on their respective held-out set and, at inference time, apply that subgroup's score quantile. 
    
    For example, we can modify the conformal approach adaptive prediction sets (APS) from \citet{NEURIPS2020_244edd7e} for the group setting:
    Let $\{(X_i, A_i, Y_i)\}_{i=1}^n$ be a set of three tuples containing a training example $X_i \in \mathrm{R}^d$, group attribute $A_i \in \{1, 2, \ldots K\}$, and target class$Y_i \in \{0, 1, \ldots, K\}$, where $K$ is the number of classes.
    The scoring function $s$ sorts the classes according to their softmax score in descending order for each example and outputs the cumulative sum until the softmax score of the true class is reached.
    Assuming all samples are drawn interchangeably from a potentially unknown distribution, then marginal coverage can be assured for each subgroup, 
    \begin{equation}
        \small
        1 - \alpha \leq P(Y_{n+1} \in C(X_{n+1} \mid A = a) \leq 1 - \alpha + \frac{1}{n + 1}
    \end{equation}
    for all $a \in A, \; \alpha \in (0, 1)$.
     
    
    \begin{algorithm}[tb]
        \small
        \caption{Group adaptive prediction sets (GAPS)}
        \label{alg:gaps}
        \textbf{Input}: 
        Held-out calibration set $\{(X, A, Y)\}^n \in \mathcal{X}^n \times \mathcal{A}^n \times \mathcal{Y}^n$, \\ 
        Miscoverage level $\alpha \in (0, 1)$, \\
        Test example $\left(x_{n+1}, a_{n+1}\right) \in \mathcal{X} \times \mathcal{A}$, \\ 
        Number of classes $k := \lvert\mathcal{Y}\rvert$, \\ 
        Prediction function $f: X \rightarrow \mathbb{R}^k$, \\
        Scoring function $s: \mathbb{R}^k \times \mathcal{Y} \rightarrow \mathbb{R} $, \\
        Sorting function $\mathrm{SORT}: \mathbb{R}^k \rightarrow \mathbb{R}^k$, \\
        Quantile function $q: \{\mathbb{R}\}^m \times [0, 1] \rightarrow \mathbb{R}$\\
        \begin{algorithmic}[1] 
            \FOR{$a \in A$}
                \STATE $s_a \leftarrow \{\ \left(f(X_i), Y_i\right) : i \in \{1, 2, \ldots, n\} \mid A_i = a\}$
                \STATE $q_a \leftarrow q(s_a, \frac{\lceil (1 - \alpha) \cdot (n + 1) \rceil}{n})$
            \ENDFOR
            \FOR{$a \in A$}
                \IF{$a = a_{n+1}$}
                    \STATE $y_{n+1} \leftarrow f(x_{n+1})$
                    \STATE $y^\prime \leftarrow \mathrm{SORT}(y_{n+1})$
                    \STATE $c_{n+1} \leftarrow \{y^\prime_j : j \in \{1, 2, \ldots, k\} \mid \sum_{j=1}^k y^\prime_j < q_a \}$
                \ENDIF
            \ENDFOR
            \RETURN $c_{n+1}$
        \end{algorithmic}
        \textbf{Output}: Conformal prediction set $c_{n+1}$ \\
    \end{algorithm}

    
\section{Experiments}
    We compare four methods: a non-conformal baseline (Naive), adaptive prediction sets (APS)~\cite{NEURIPS2020_244edd7e}, regularized adaptive prediction sets (RAPS)~\cite{angelopoulos2021uncertainty}, group adaptive prediction sets (GAPS), and group regularized adaptive prediction sets (GRAPS).
    In our experiments, we investigate the following questions:
    \begin{enumerate}
        \item How do different methods compare in \textit{rule-in} and \textit{rule-out} use cases for malignant skin lesion classification? 
        \item How do group conformal predictors compare with aggregate conformal predictors conditioned on skin type as the group attribute?
        \item What is the relationship between prediction set size and epistemic uncertainty?
    \end{enumerate}
    
    \subsection{Skin Lesion Dataset}
        \begin{figure}[t]
        \centering
        \includegraphics[width=0.99\columnwidth]{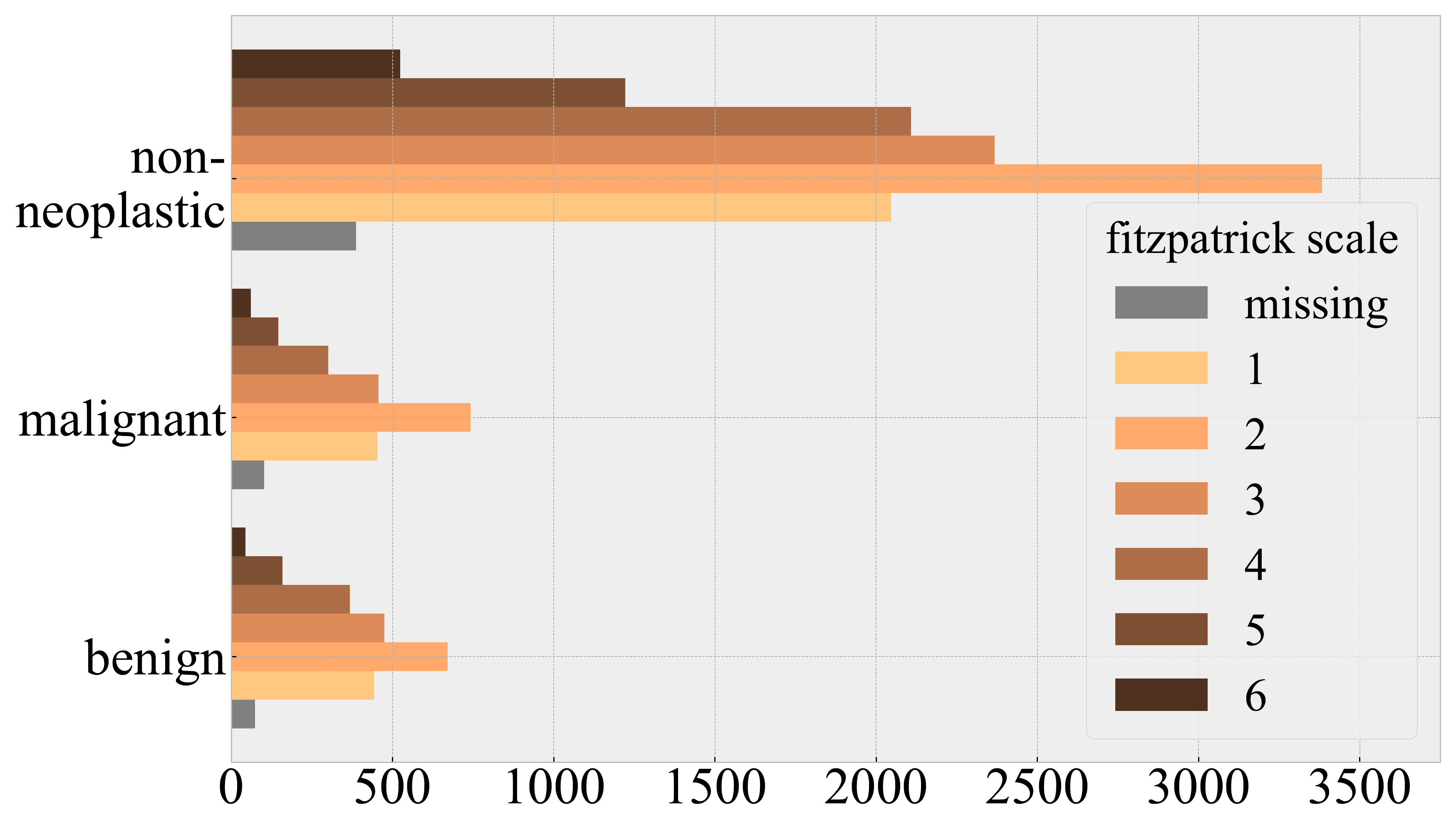}
        \caption{Distribution of skin conditions by Fitzpatrick skin type and categorization of the 114 different lesions into one of three broad categories: non-neoplastic, malignant, or benign.}
        \label{fig:fitz-dist}
        \end{figure}
        
        \begin{table}[t]
        \small
        \centering
        \begin{tabular}{cccc}
        \toprule
        \textbf{skin type} & \bf{non-neoplastic} & \bf{benign} & \bf{malignant} \\
        \midrule
        1 & 69.6\% & 15.0\% & 15.4\% \\
        2 & 70.5\% & 14.0\% & 15.5\% \\
        3 & 71.8\% & 14.4\% & 13.8\% \\
        4 & 75.9\% & 13.2\% & 10.8\% \\
        5 & 80.0\% & 10.4\% & 9.6\% \\
        6 & 83.4\% & 7.0\% & 9.6\% \\
        missing & 68.9\% & 13.0\% & 18.1\% \\
        \midrule
        total & 72.8\% & 13.5\% & 13.7\% \\
        \bottomrule
        \end{tabular}
        \caption{Percentage of skin conditions in each of the three broad dermatological categories for different Fitzpatrick skin type subgroups.}
        \label{tab:disease-prevalence}
        \end{table}
        
        For our experiments, we use the Fitzpatrick17k dataset, which aggregates $16,577$ photography images from two dermatology atlases~\cite{Groh2021EvaluatingDN}. 
        This dataset contains a hierarchical labeling scheme of $114$ different skin conditions, which are further aggregated into three dermatological categories: non-neoplastic lesions (12,080), benign lesions (2,234), and malignant lesions (2,253).
        Additionally, each image is labeled with a Fitzpatrick skin type -- an ordinal 6-point scale that estimates the amount of melanin pigment in the skin.
        The disease distribution for each skin type is shown in Figure~\ref{fig:fitz-dist} and Table~\ref{tab:disease-prevalence}.
        
        As the prevalence of skin conditions such as melanoma is known to differ between demographics such as race and ethnicity, we treat the Fitzpatrick skin type as the group attribute in our experiments.
        We also treat the $11$ malignant skin conditions as ``critical'' conditions and the other $103$ conditions as ``non-critical'' for the purposes of our two conformal prediction use-cases.
        For the ``rule-in'' scenario, only critical cases with one of the $11$ malignant skin condition were considered, with a prediction was considered correct if the true class was contained in the prediction set.
        For the ``rule-out'' scenario, only non-critical cases were considered, with a prediction was considered correct if none of the $11$ malignant conditions were contained in the prediction set.
        
    \subsection{Evaluation metrics}
        We consider our methods with two evaluation metrics: \textit{marginal coverage} and \textit{set size}.
        Given some miscoverage parameter, $\alpha$, marginal coverage is defined as the number of times the true class is contained in the prediction set, averaged over all examples.
        
        Set size is the average number of elements in the prediction set.
        To quantify difference between subgroups, we also define \textit{coverage disparity} to be the average pairwise difference in marginal coverage
        \begin{equation} \label{eq:cov-disparity}
        \small
        \frac{1}{\mid A \mid} \sum_{a, b \in \binom{A}{2}} \big\lvert \text{coverage}_{A=a} - \text{coverage}_{A=b} \big\rvert,
        \end{equation}
        where $\binom{A}{2}$ is all pairwise combinations of subgroups.
        
        We define \textit{set size disparity} in a similar manner 
        \begin{equation} \label{eq:card-disparity}
        \small
        \frac{1}{\mid A \mid} \sum_{a, b \in \binom{A}{2}} \big\lvert \text{set size}_{A=a} - \text{set size}_{A=b} \big\rvert.
        \end{equation}
        
    \subsection{Implementation Details}
        We use a ResNet-18~\cite{DBLP:journals/corr/XieGDTH16}, pretrained on ImageNet~\cite{imagenet_cvpr09}, as our deep learning image classifier and average across five different initializations.
        We use Monte Carlo dropout of $0.1$ and $T=30$ to estimate maximum softmax probability~\cite{hendrycks17baseline} and predictive entropy (Equation~\ref{eq:entropy}) with softmax scores calibrated using Platt scaling (Equation~\ref{eq:platt}).
        For our non-conformal baseline, we form prediction sets by sorting softmax scores and taking classes until the $1 - \alpha$ threshold.
        All models were trained with cross-entropy for $100$ epochs, early stopping of $20$ epochs, a learning rate of $0.0001$, and a batch size of $16$. 
        All experiments were implemented using the PyTorch framework~\cite{NEURIPS2019_9015} and trained on a Nvidia A100 GPU.
        Code and data to reproduce our experiments is made available here: https://github.com/clu5/AAAI22.

\section{Results}

    \begin{figure*}[t]
    \centering
    \includegraphics[width=0.99\textwidth]{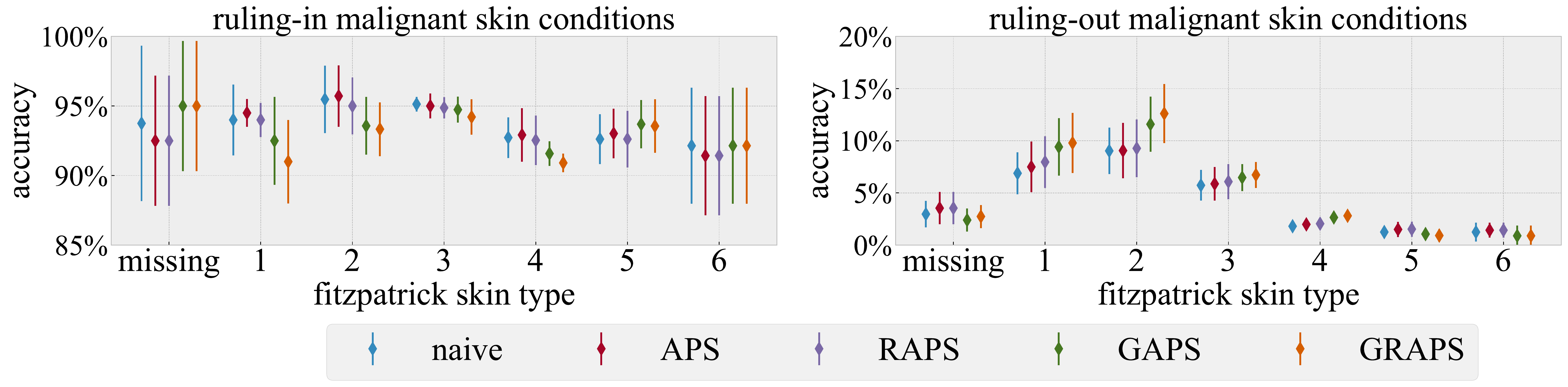}
    \caption{Subgroup accuracy of different prediction set methods at $\alpha=0.1$ for ruling-in and ruling-out use-cases.}
    \label{fig:fitz-use-case}
    \end{figure*}
    
    \begin{figure*}[t]
    \centering
    \includegraphics[width=0.95\textwidth]{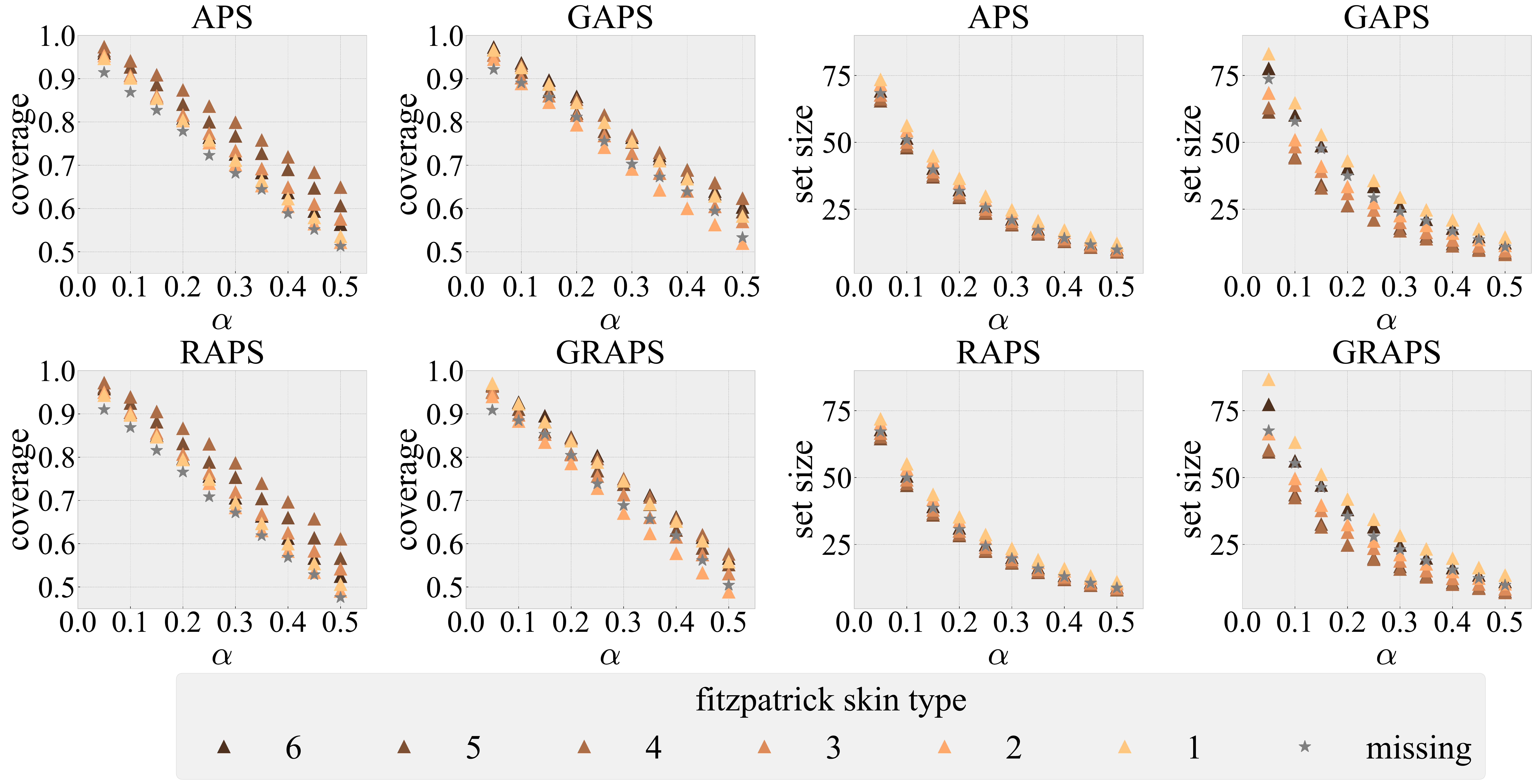} 
    \caption{Comparing coverage and set size of group conformal methods (GAPS and GRAPS) and aggregate conformal methods (APS and RAPS) at different $\alpha$ values for skin lesion classification; colors denote Fitzpatrick skin types; gray star represents the missing skin type.}
    \label{fig:fitz-coverage}
    \end{figure*}
    
    \begin{figure*}[t]
    \centering
    \includegraphics[width=0.99\textwidth]{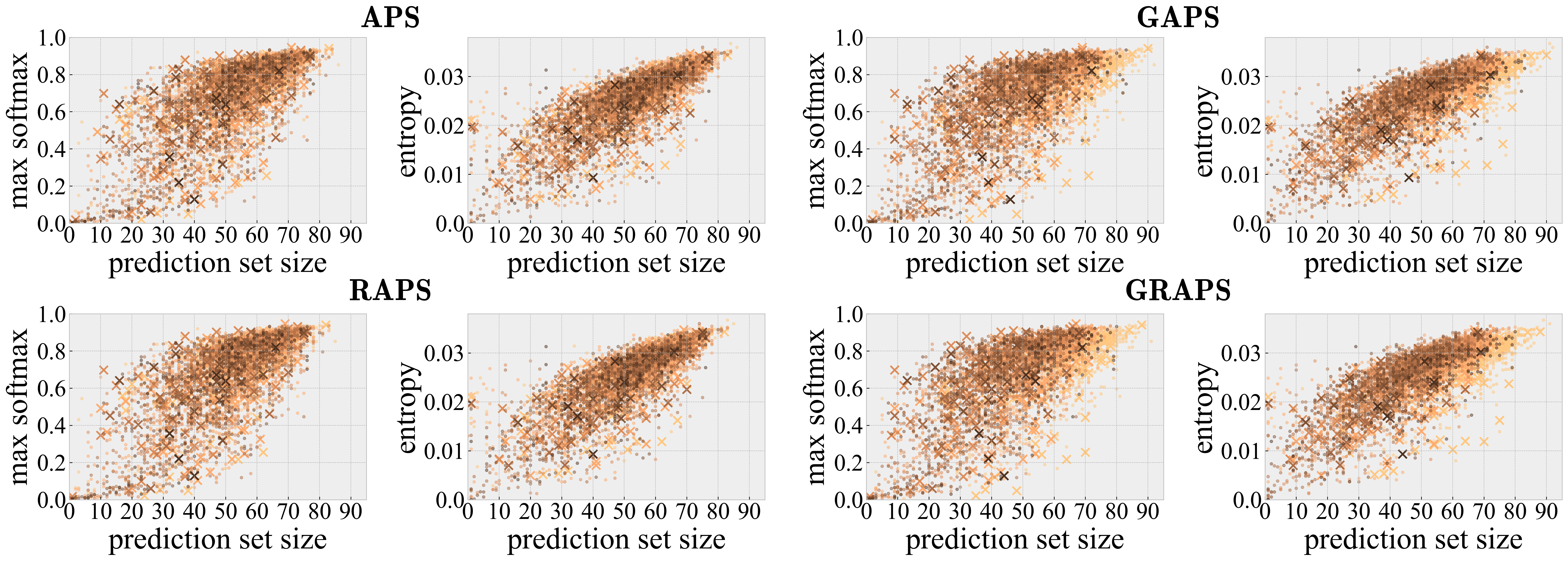}
    \caption{Relationship between conformal uncertainty and two epistemic uncertainty measures at $\alpha=0.1$; colors correspond to different Fitzpatrick skin types; $\times$ denotes a malignant skin condition and $\circ$ denotes a non-malignant skin condition.}
    \label{fig:epistemic-uncertainty}
    \end{figure*}
    
    For our two use-cases, we find all four methods of prediction sets to perform similarly at $\alpha=0.1$ in terms of accuracy and set size at a subgroup level (Figure~\ref{fig:fitz-use-case}). 
    Group conformal methods (GAPS and GRAPS) in performing slightly better at ruling-out malignant skin conditions and worse at ruling-out malignant skin conditions than other methods for skin types 1 and 2 (the lightest skin tones).
    
    Next, we compare performance between Fitzpatrick skin types in terms of marginal coverage and set size between aggregate conformal methods (APS and RAPS) and their group variants (GAPS and GRAPS) in Figure~\ref{fig:fitz-coverage}.
    We observe lower subgroup coverage disparity 
    as well as higher set size disparity with GAPS and GRAPS compared to APS and RAPS (Table~\ref{tab:disparity}).
    
    Finally, we consider the relationship between set size and epistemic uncertainty, measured by maximum softmax probability and predictive entropy, and find a strong correlation between set size and epistemic uncertainty.
    Looking at Figure~\ref{fig:epistemic-uncertainty}, we can see that the group conformal methods show increased separation between Fitzpatrick subgroups -- particularly for skin type 1 and skin type 6 -- compared to the naive baseline and aggregate conformal methods.
    This separation is quantified in Table~\ref{tab:uncertainty-correlation}, which shows that GAPS and GRAPS have lower Spearman correlation than Naive, APS, and RAPS across a range of $\alpha$.

    \begin{table}[t]
    \small
    \centering
    \begin{sc}
    \begin{tabular}{c|rccccc}
    \toprule
    & \bf{method} & $0.1$ & $0.2$ & $0.3$ & $0.4$ & $0.5$ \\
    \midrule
    \multirow{5}{*}{\rotatebox[origin=c]{90}{\bf{coverage}}}
    & naive & 0.026 & 0.040 & 0.059 & 0.070 & 0.080 \\
    & APS   & 0.026 & 0.035 & 0.048 & 0.056 & 0.059 \\
    & RAPS  & 0.026 & 0.037 & 0.049 & 0.054 & 0.057 \\
    & GAPS  & 0.024 & 0.029 & 0.037 & 0.038 & 0.045 \\
    & GRAPS & 0.022 & 0.028 & 0.037 & 0.035 & 0.039 \\
    \midrule
    \multirow{5}{*}{\rotatebox[origin=c]{90}{\bf{set size}}}
    & naive &  3.9 & 2.2 & 1.2 & 0.6 & 0.2 \\
    & APS   &  3.5 & 3.1 & 2.4 & 1.9 & 1.4 \\
    & RAPS  &  3.6 & 3.0 & 2.4 & 1.8 & 1.3 \\
    & GAPS  &  9.7 & 8.1 & 5.6 & 4.2 & 2.8 \\
    & GRAPS &  9.1 & 8.0 & 5.5 & 4.2 & 2.7 \\
    \bottomrule
    \end{tabular}
    \end{sc}
    \caption{Subgroup disparity in coverage and set size between Fitzpatrick skin types at five different levels of $\alpha$.}
    \label{tab:disparity}
    \end{table}
    
    \begin{table}[t]
    \small
    \centering
    \begin{sc}
    \begin{tabular}{c|rccccc}
    \toprule
    & \bf{method} & $0.1$ & $0.2$ & $0.3$ & $0.4$ & $0.5$ \\
    \midrule
    \multirow{5}{*}{\rotatebox[origin=c]{90}{\bf{softmax}}}
    & naive & 0.71 & 0.75 & 0.78 & 0.75 & 0.55 \\
    & APS   & 0.69 & 0.72 & 0.75 & 0.78 & 0.82 \\
    & RAPS  & 0.69 & 0.73 & 0.75 & 0.79 & 0.82 \\
    & GAPS  & 0.66 & 0.70 & 0.74 & 0.76 & 0.80 \\
    & GRAPS & 0.66 & 0.70 & 0.74 & 0.76 & 0.80 \\
    \midrule
    \multirow{5}{*}{\rotatebox[origin=c]{90}{\bf{entropy}}}
    & naive & 0.82 & 0.84 & 0.80 & 0.66 & 0.34 \\
    & APS   & 0.81 & 0.83 & 0.85 & 0.87 & 0.87 \\
    & RAPS  & 0.81 & 0.84 & 0.85 & 0.87 & 0.87 \\
    & GAPS  & 0.77 & 0.81 & 0.84 & 0.84 & 0.86 \\
    & GRAPS & 0.77 & 0.81 & 0.84 & 0.84 & 0.86 \\
    \bottomrule
    \end{tabular}
    \end{sc}
    \caption{Spearman correlation between set size and epistemic uncertainty (maximum softmax probability and predictive entropy) at five different values of $\alpha$.}
    \label{tab:uncertainty-correlation}
    \end{table}
    
\section{Discussion}
    In this paper, we explored might be conformal predictors integrated into medical imaging workflows. 
    Our interviews suggested several ways in which conformal predictions could allow for greater transparency in medical AI, such as the identification of possible clinical mimics in a manner similar to differential diagnoses.
    We also formulated two general use-cases for conformal predictions for clinical decision-making.
    
    To better calibrate conformal predictors to subgroup differences, we modified two methods of conformal predictions (APS and RAPS) to guarantee equalized coverage for known demographic attributes and performed experiments on a heterogeneous skin lesion dataset with Fitzpatrick skin type to evaluate group conformal predictors.
    Results from our experiments demonstrated that group conformal predictors have lower coverage disparity than aggregate conformal methods and show greater variability in set size at a subgroup level, reflecting underlying differences between different skin types. 
    Specifically, skin types 1 and 6 have larger prediction set sizes, indicating more uncertainty.
    These differences may be due to the fact that lesions of patients with lighter skin tones have a higher rate of malignancy (Table~\ref{tab:disease-prevalence}) while patients with the skin type 6 are underrepresented in the dataset, comprising only about 4\% of the total dataset.
    
    
    Further comparing group conformal uncertainty against aggregate conformal uncertainty, we observed that group conformal methods are less correlated with epistemic uncertainty measures, thus indicating more adaptiveness to subgroup differences across task difficulty levels as measured by maximum softmax probability and predictive entropy.
    This observation agrees with visual inspection of the scatter plots in Figure~\ref{fig:epistemic-uncertainty} that show increased separation for both malignant and non-malignant skin conditions for the lightest and darkest skin tones.
    Therefore, we conclude that group conformal methods can better describe subgroup uncertainty than regular conformal methods when subgroups come from different data distributions.
    
\section{Conclusion}
    Fair conformal predictors have the potential to increase clinician trust in AI models by providing meaningful notions of uncertainty across clinically relevant sub-populations.
    These distribution-free methods complement existing deep learning models and require no modifications to existing training procedures. 
    Based on our experiments, we find group conformal predictors to be a promising and generally applicable approach to increasing clinical usability and trustworthiness in medical AI.
    We hope this work promotes further work into group conformal predictors for clinical applications in healthcare.
    
\section*{Acknowledgements}
    The authors would like to thank Christopher Bridge, Bernardo Bizzo, Stuart Pomerantz, and James Hillis for their help and support with this work.

\bibliography{aaai22.bib}

\appendix

    
\end{document}